# Discussion of Parameters Setting for A Distributed Probabilistic Modeling Algorithm

Mengshuo Jia. *Student Member, IEEE,* Chen Shen, *Senior Member*, IEEE, *Zhiwen Wang, Student Member, IEEE*

***Abstract*—This manuscript provides additional case analysis for the parameters setting of the distributed probabilistic modeling algorithm for the aggregated wind power forecast error.**

***Index Terms*—additional case analysis, parameters setting**

## I. Introduction

This manuscript provides additional case analysis for the parameters setting of the proposed distributed probabilistic modeling algorithm for the aggregated wind power forecast error. These case analysis aim to discuss the parameters setting of the choice of key newly-built wind farms (NWFs), the order of GMM, and the distance threshold for neighbor NWFs.

The rest of the manuscript is organized as follows. In Section II, the discussion of the physical meaning of the k-coreness is given. Meanwhile, the comparisons between different choice of key NWFs are illustrated. In Section III, the influence of different GMM's order on the estimation effect of the proposed algorithm is discussed. In Section IV, estimation results under different thresholds for neighbor NWFs are investigated.

## II. Discussion of K-coreness

### *A. The physical meaning of the k-coreness*

First, as a concept from graph theory, the application of k-coreness does not involve the characteristics of wind or wind power, but simply involves information exchange between adjacent nodes in a communication network.

Second, the information exchange between adjacent NWFs can be viewed as information spreading in a communication network, and k-coreness can be used to identify influential spreaders [1]. If $k_s$ denotes the value of k-coreness, then the physical meaning of k-coreness is as follows, quoting from [1]:

*"In conclusion, the nodes with the largest $k_s$ values consistently a) are infecting larger parts of the network b) are infected more frequently."*

Therefore, whether the information is transmitted or received, a node with a larger k-coreness value is more comprehensive than a node with a smaller k-coreness value, and the larger one can make a more accurate estimation of the global statistics. To verify this conclusion, based on the 30-day training data set, we construct the joint PDFs by the proposed distributed MAP (DMAP) estimation, which is not involved the VN. Since the PDF built by the centralized MAP estimation can be viewed as a benchmark, the RMSEs between the PDF constructed by the NWFs and the benchmark can be calculated. Then the average RMSEs of the NWFs with the same k-coreness value are given in Table I.

TABLE I
Average RMSE For Different K-coreness

| K-coreness Value | Average RMSE |
|---|---|
| 4 | 0.0015 |
| 2 | 0.0048 |
| 1 | 0.0065 |

It can be seen from Table RI that the larger the k-coreness value, the smaller the average RMSE and the more accurate the estimation. Because the purpose of introducing VN is to obtain the most accurate global estimates, we link VN to NWFs with larger k-coreness values.

It should be noted that there may be multiple nodes with the same k-coreness value. For example, node 2, node 4, node 5, node 7 and node 9 have the same k-coreness value. However, the RMSEs of these nodes are slightly different. Therefore, those RMSEs can be used to order the nodes with the same k-coreness value.

### *B. Comparison between the results by choice of k-coreness and random choice of other nodes*

We first consider two extreme situations: 1) setting the top 30% of nodes by k-coreness (i.e., node 2, node 4 and node 5) as key nodes; 2) set the last 30% of nodes by k-coreness (i.e., node 3, node 6 and node 8) as key nodes. Then, we randomly choose another 3 groups of key nodes. The number of groups of key nodes is shown in Table II.

TABLE II
Group For Different Choice of Key Nodes

| Group | 1 | 2 | 3 | 4 | 5 |
|---|---|---|---|---|---|
| Key Nodes | 2, 4, 5 | 2, 5, 7 | 1, 3, 10 | 1, 6, 3 | 8, 6, 3 |

Under varied groups of key nodes, we construct the joint PDFs by the proposed algorithm. Since the PDF built by the centralized MAP estimation can be viewed as a benchmark, the RMSEs between the PDF constructed by VN and the benchmark are given in Fig. 1.

In Fig. 1, the estimation of the first group is much better than that of the fifth group. For the random groups of key nodes, the corresponding RMSEs are between the first group and the fifth group. In addition, the fewer nodes with low k-coreness values in the selection of key nodes, the more accurate the estimation of VN. For example, the fifth group has three nodes with the lowest k-coreness value, the fourth group has two nodes with the lowest k-coreness value, and the third group has one. It can be seen that the estimation effect of VN gradually improves from the fifth group to the third group. Conversely, the fewer nodes with high k-coreness in the group, the less accurate the estimation of VN, i.e., from the first group to the second group, the estimation of VN declines.

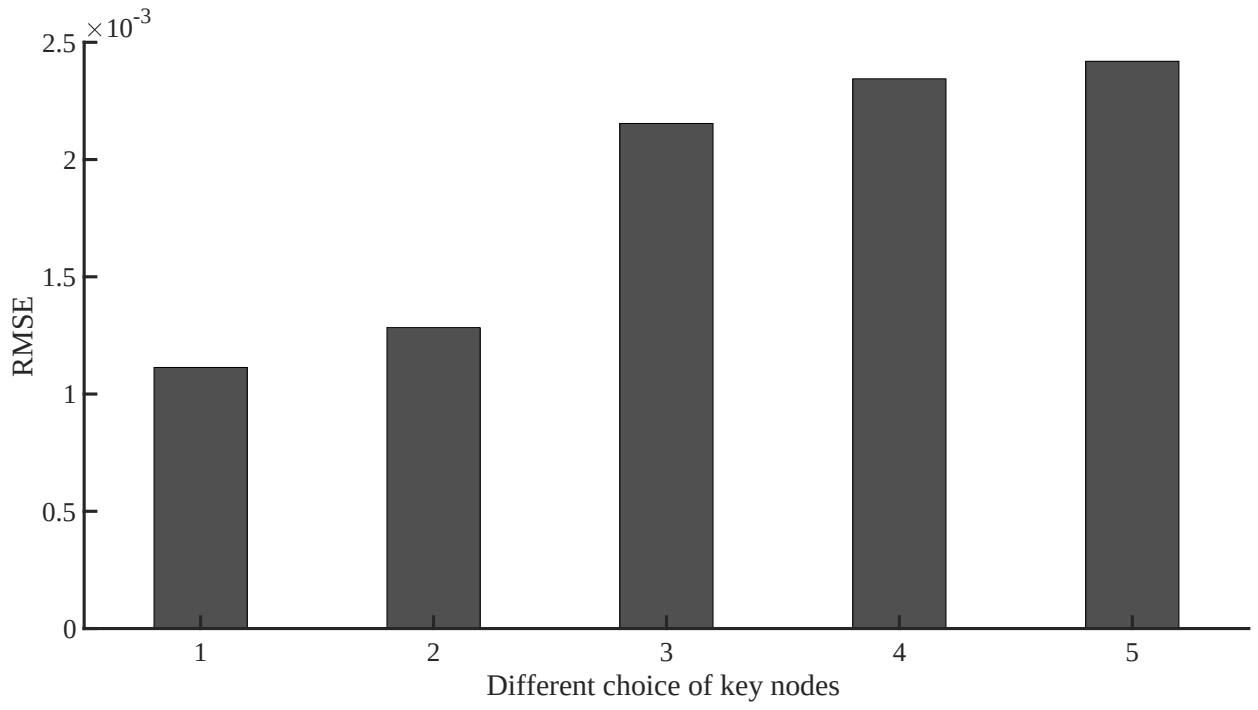

Figure. 1. RMSE for different choice of key nodes.

### C. *Comparison between the number of key nodes*

We gradually increase the percentage of k-coreness from top 10% to top 50% to set the key nodes. The key nodes for various percentages are given in Table RIV. Since the PDF built by the centralized MAP estimation can be seen as a benchmark, the RMSEs between the PDF constructed by VN and the benchmark under various key nodes are illustrated in Fig. 2.

TABLE III
KEY NODES FOR DIFFERENT PERCENTAGE OF K-CORENESS

| Percentage | 10% | 20% | 30% | 40% | 50% |
|---|---|---|---|---|---|
| Key Nodes | 5 | 5,4 | 5,4,2 | 5,4,2,7 | 5,4,2,7,9 |

The more nodes connected to VN, the more information VN can acquire and the better the estimation VN can obtain. However, when the number of connected nodes reaches a certain level, the estimation accuracy of VN tends to become saturated. As shown in Fig. R8, from the top 10% to 30%, RMSE decreases obviously. However, from the top 30% to 50%, RMSE tends to be stable. Therefore, the key nodes are set as the nodes with top 30% k-coreness.

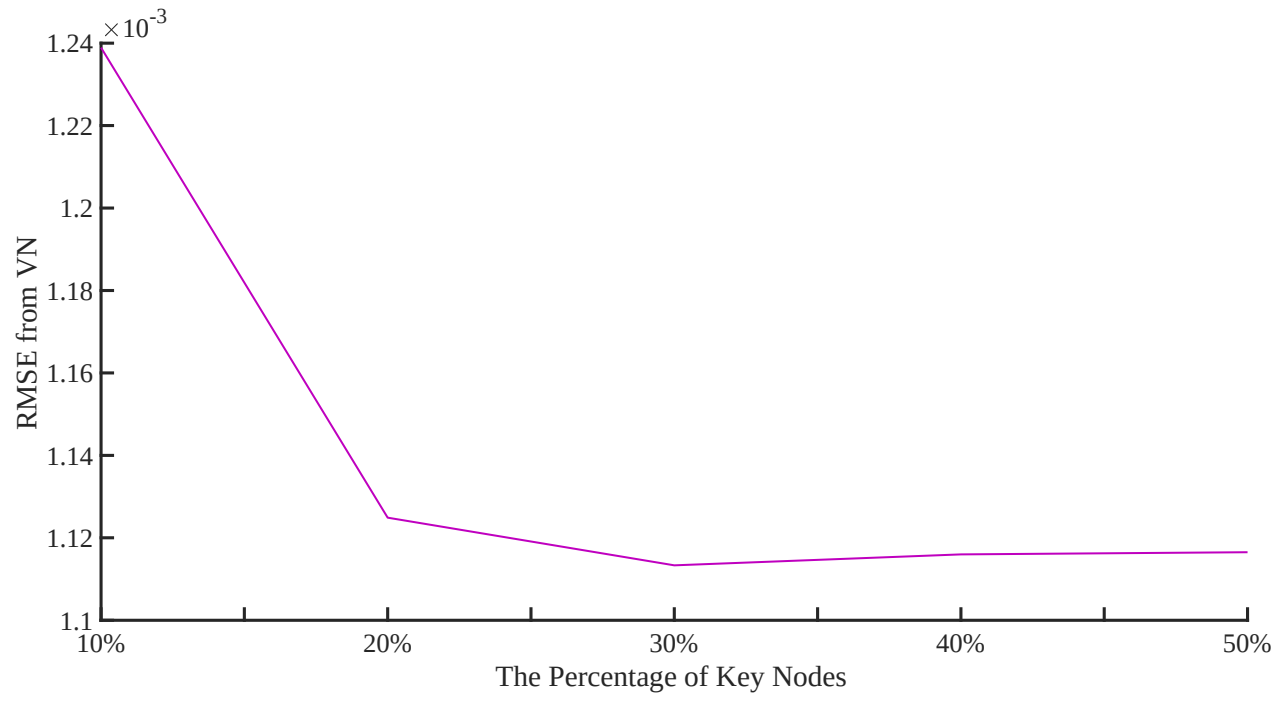

Figure. 2. The RMSE from VN for different percentage of key nodes.

## III. DISCUSSION OF THE ORDER OF GMM

The number of GMM components is also known as the GMM order. For the proposed algorithm and the centralized MAP estimation, they have the same GMM order and the order is set via case experiments. First, a 100-day testing data set is used to formulate the empirical distribution. Then, based on the 30-day training data set, we gradually increase the order from 1 to 30 and record the RMSEs between the PDF constructed by VN and the empirical distribution. The results are given in Fig. R4. When the order is too small, it is difficult to fit the empirical distribution, i.e., RMSE is high. As the order increases, RMSE gradually decreases and tends to be stable. Since RMSE is the smallest when the order is 20, we choose this GMM order for the proposed algorithm.

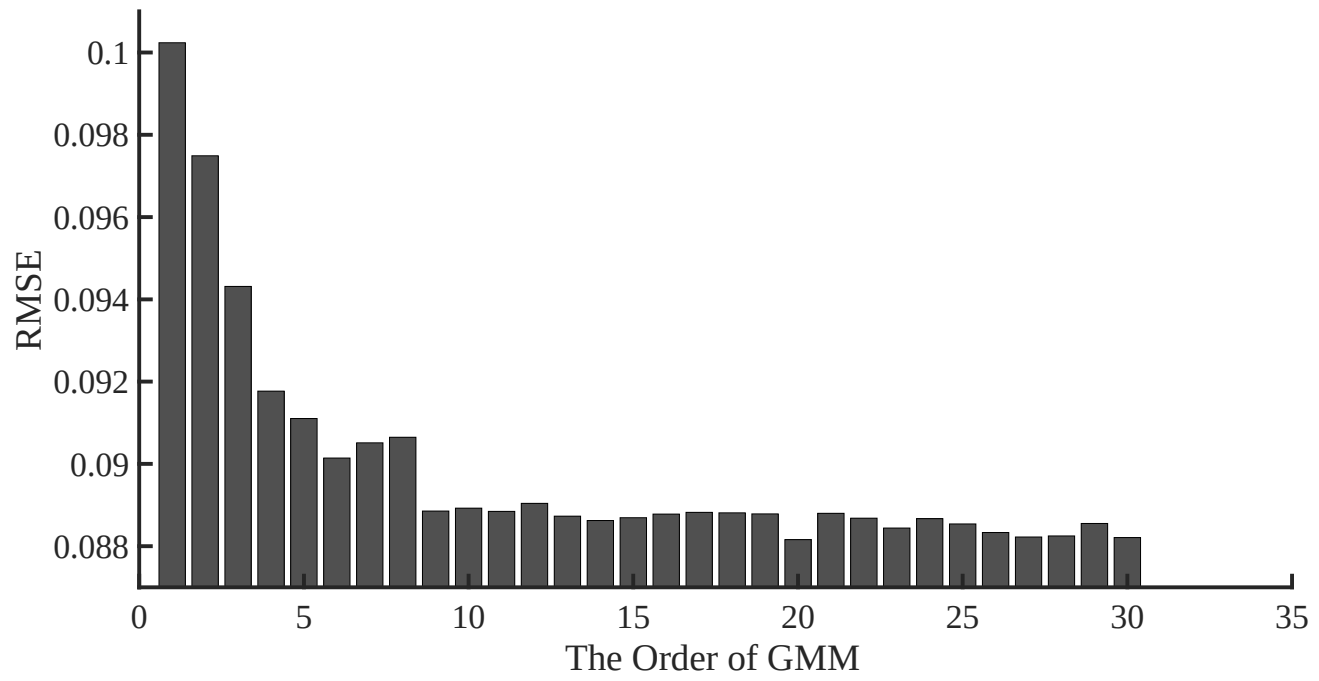

Figure. 3. RMSE for different order of GMM.

Moreover, we need to compare the proposed algorithm and the centralized EM algorithm. Therefore, it is necessary to investigate whether 20 components are suitable for the centralized EM algorithm. As a kind of maximum likelihood estimation method, there are many evaluation indices based on the likelihood function to guide the order setting of GMM for the centralized EM algorithm. The most popular one is the Akaike information criterion (AIC) [2]. Based on Kullback–Leibler divergence, AIC can be used to judge the degree of information loss between the training data set and trained GMMs of various orders [3]. The smaller the AIC, the smaller the information loss.

Based on the 30-day training data set, we gradually increase the order from 1 to 30 and record the AIC values. The results are illustrated in Fig. 4, where 23 components correspond to the minimum AIC, and 20 components correspond to the second smallest AIC.

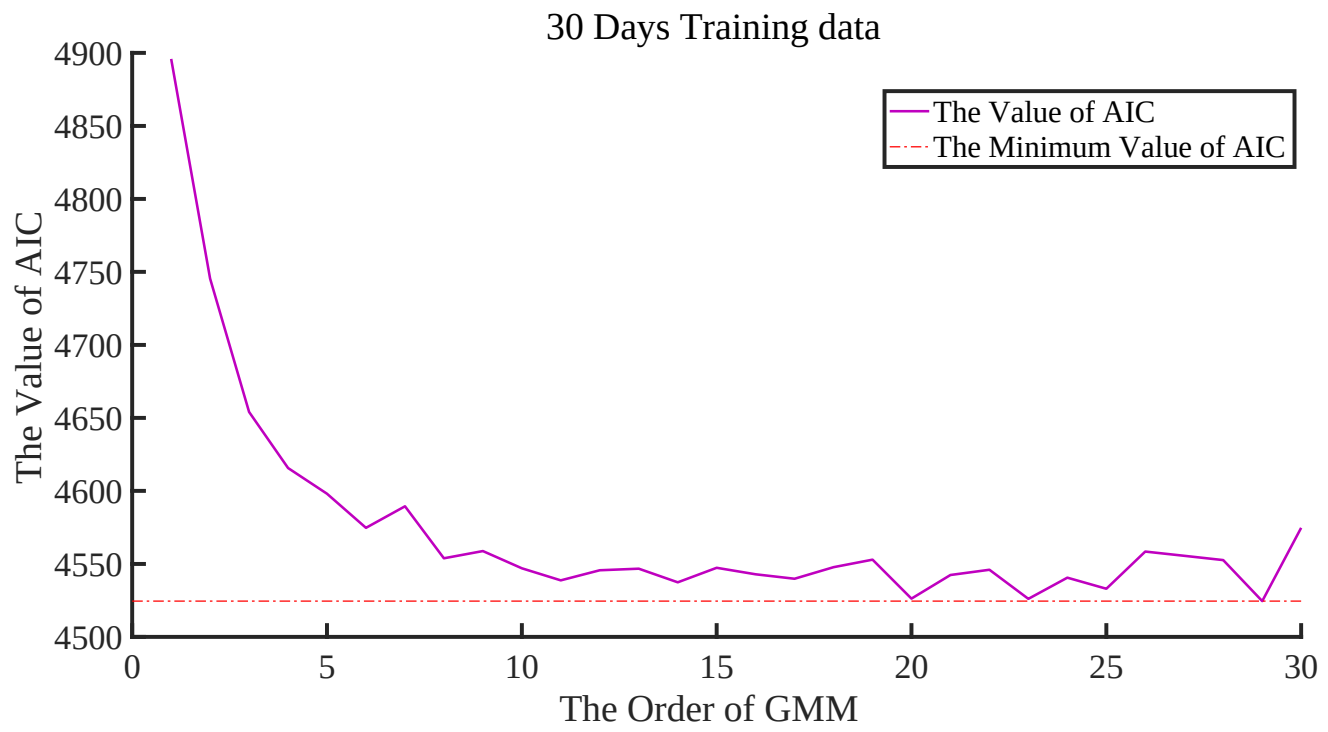

Figure. 4. The value of AIC for different order of GMM.

A further comparison of the two orders is provided in Table IV. Although the AIC of 20 components is higher than that of 23 components, the difference is only 0.0033%. Meanwhile, the RMSE between the PDF constructed by VN and the empirical distribution only differ by 1.18%. Moreover, the computation time of 20 components is 17.07% lower than that of 23 components. Therefore, 20 components are considered appropriate for the centralized EM algorithm with the consideration of RMSE and computational cost.

TABLE IV
COMPARISON BETWEEN ORDER 20 AND ORDER 23

| | Order 23 | Order 20 | Decreased by order 20 |
|---|---|---|---|
| AIC | 4526.031 | 4526.182 | -0.0033% |
| RMSE | 0.085 | 0.086 | -1.18% |
| Computation Time | 1.23seconds | 1.02seconds | 17.07% |

Notably, even if the centralized EM algorithm trains a GMM with 23 components while the proposed algorithm trains a GMM with 20 components, the corresponding conclusions can still be consistent with the conclusions in the original manuscript, as shown in Fig. 5 and Fig. 6. In Fig. 5, the RMSE of the PDF constructed by the centralized EM algorithm with 23 components are still higher than that by the proposed algorithm with 20 components, which is consistent with Fig. 8 in the manuscript. In Fig. 6, the overfitting of the centralized EM algorithm with 23 components is still obvious, which is consistent with Fig. 9 in the manuscript.

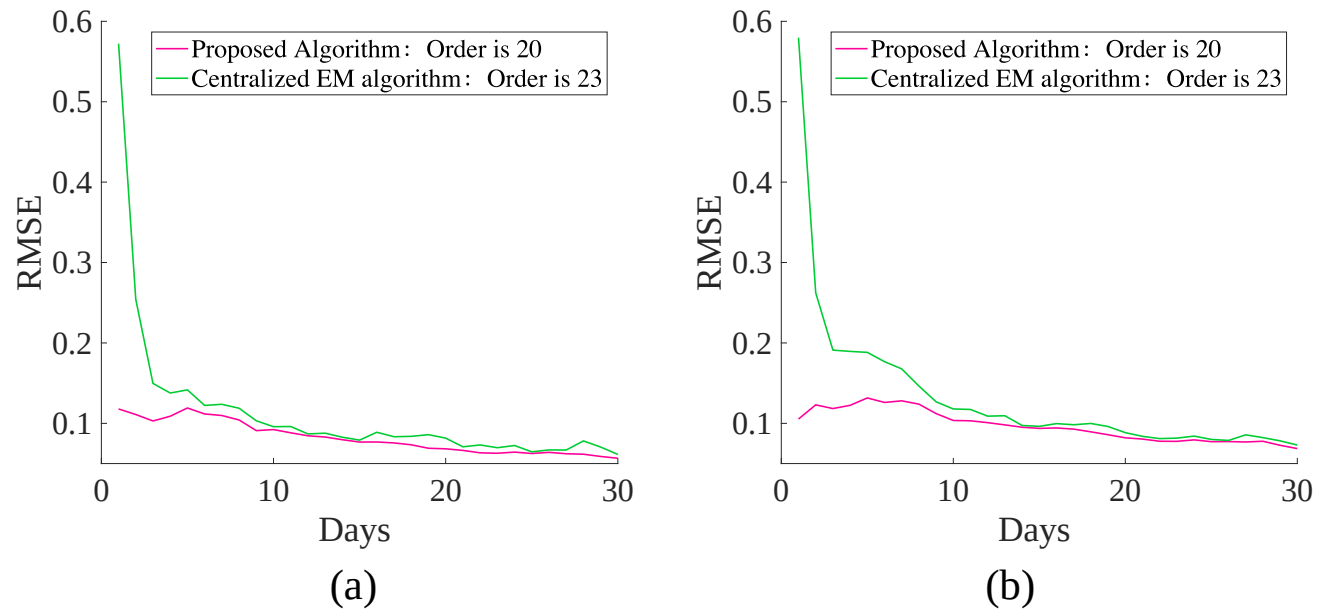

Figure. 5. The RMSEs between the empirical marginal distribution of the aggregated AWO and the corresponding marginal PDF constructed by the two algorithms (b) The RMSEs between the empirical marginal distribution of the aggregated FWO and the corresponding marginal PDF constructed by the two algorithms

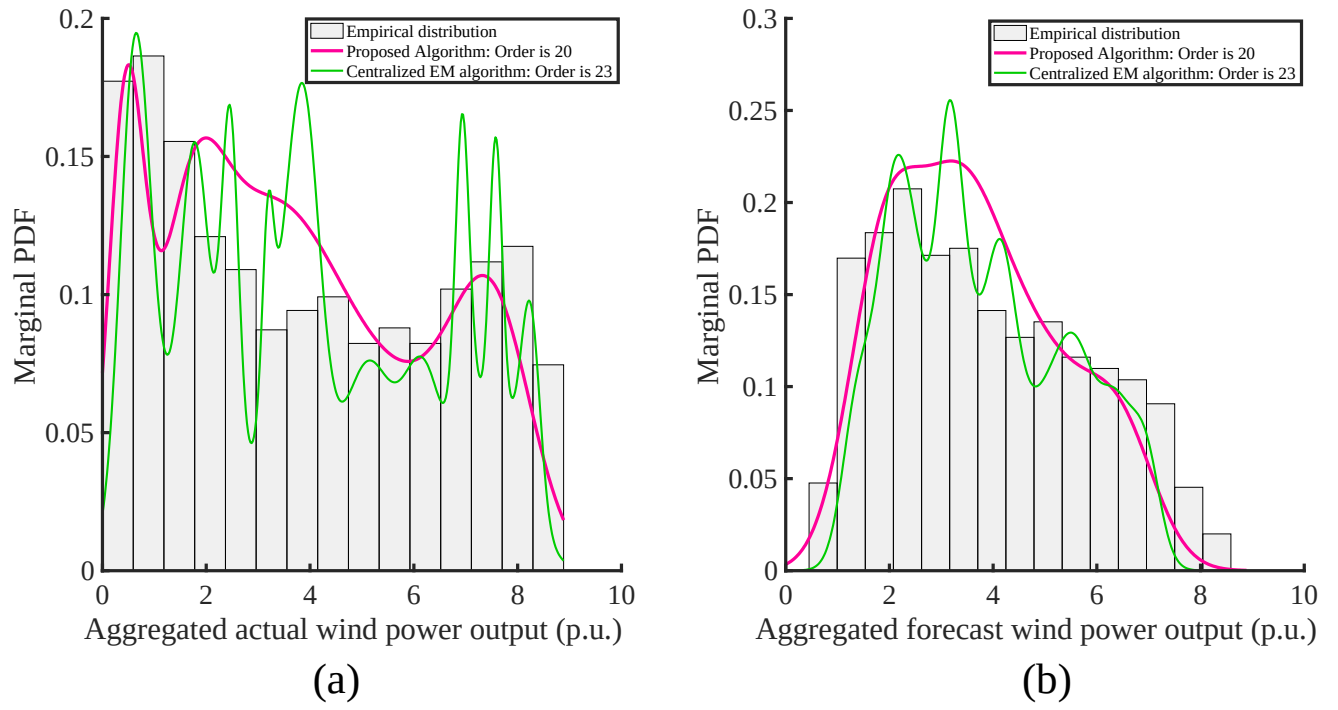

Figure. 6. The RMSEs between the empirical marginal distribution of the aggregated AWO and the corresponding marginal PDF constructed by the two algorithms (b) The RMSEs between the empirical marginal distribution of the aggregated FWO and the corresponding marginal PDF constructed by the two algorithms

In the end, in order to maintain the uniformity of the comparison form, both the centralized EM algorithm and the proposed algorithm train a GMM with 20 components.

## IV. The Discussion of The Distance Threshold

We gradually increase the threshold from 3 km to 8.5 km. When the threshold reaches 4 km, the whole communication network is connected, which is a basic premise to ensure that the proposed algorithm can achieve a global consensus. When the threshold reaches 8.5 km, all nodes in the communication network are connected to other nodes. The communication networks for various thresholds are illustrated in Fig. 7.

Since the PDF built by the centralized MAP estimation can be viewed as a benchmark, the average RMSEs between the benchmark and the PDF constructed by all NWFs and VN are shown in Fig. 8 for various thresholds; the total lengths of all the communication lines, which indicates the cost of communication construction, are also given in Fig. 8.

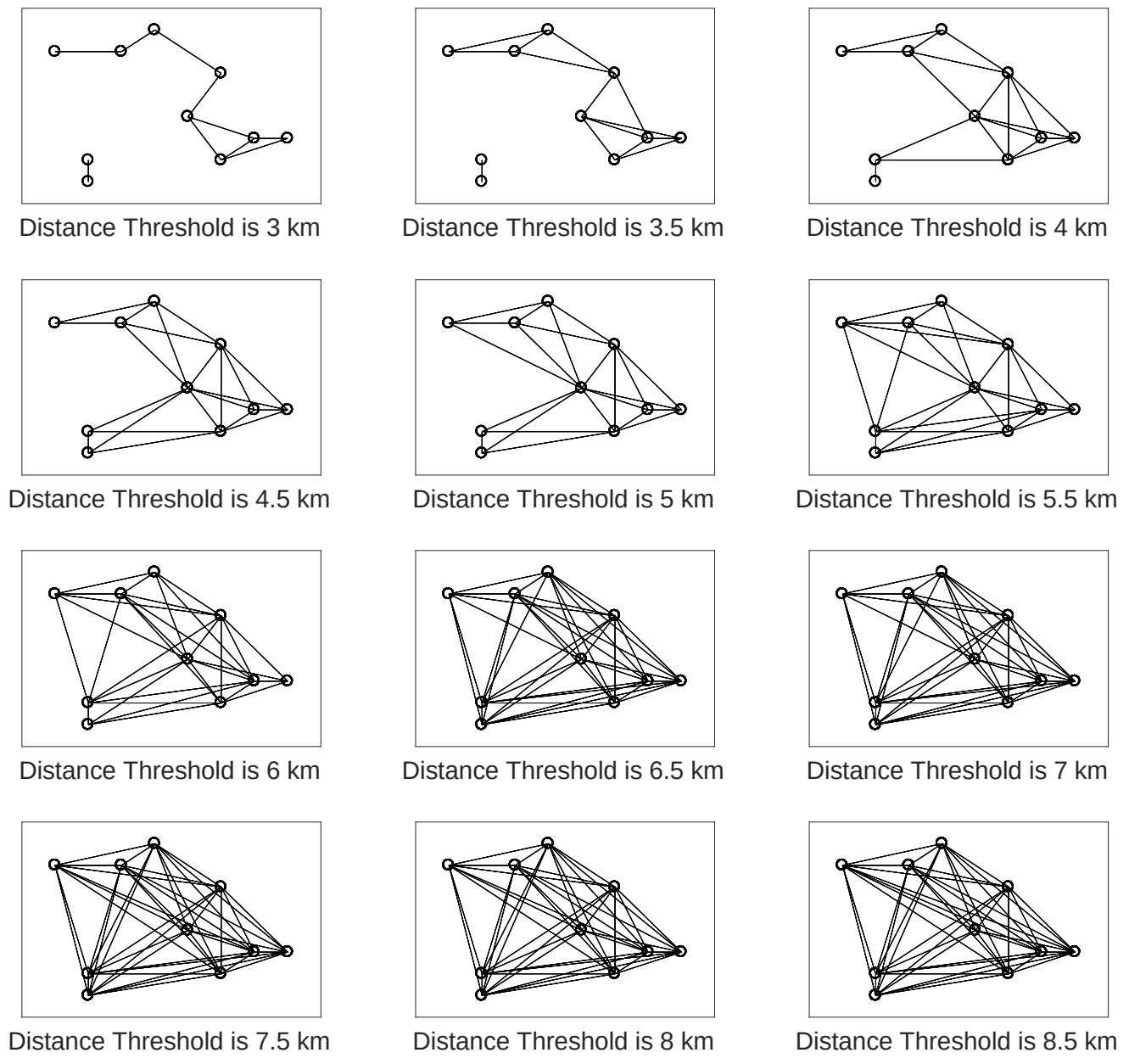

Figure. 7. The communication network for different distance threshold

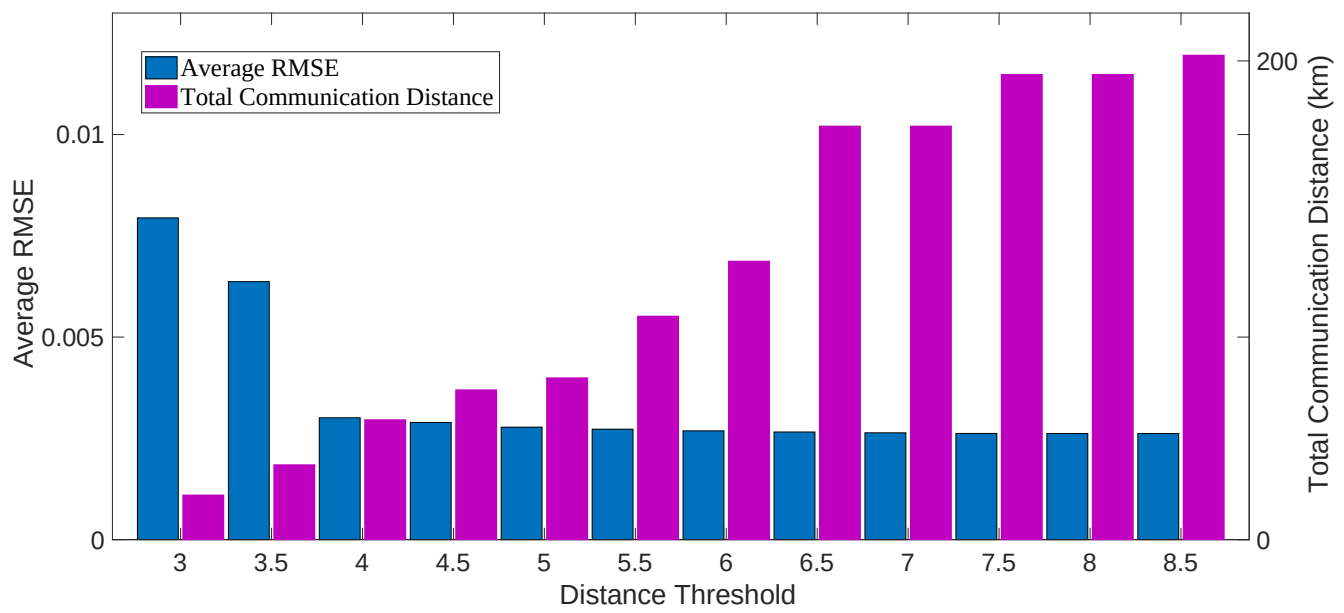

Figure. 8. The average RMSE and total communication distance for all nodes

In Fig. 8, the average RMSE decreases sharply when the threshold varies from 3 km to 4 km. Once network connectivity is achieved, the average RMSE is much lower than those when the network is disconnected. However, when the proposed algorithm can achieve a global consensus, the average RMSE tends to be stable when the threshold varies from 4 km to 8.5 km. Although the average RMSE is still decreasing after the threshold reaches 4 km, the reduction is very limited compared to that when the network is disconnected. Moreover, the larger the threshold, the longer the length of all the communication lines, which indicates higher cost for communication construction.

Therefore, by considering both the estimation effect and construction costs, we set the threshold for a neighbor node to 4 km.